\documentclass[final,5p,times,twocolumn,authoryear]{elsarticle}

\usepackage{amsmath,amssymb,amsfonts}
\usepackage{booktabs}

\newcommand{\eps}{\varepsilon}
\newcommand{\Zthr}{Z_3}
\newcommand{\ZZ}{Z_3\!\times\!Z_3}
\newcommand{\Tsix}{T^6\!/(\ZZ)}

\newcommand{\VEV}[1]{\langle #1 \rangle}

\newcommand{\Kah}{K\"ahler}

\journal{Nuclear Physics B}

\begin{document}

\begin{frontmatter}

\title{The exact column texture: tree-level Yukawa universality in heterotic $\ZZ$ orbifolds}

\author[first]{Navid Ardakanian}
\ead{n.ardakanian@gmail.com}
\affiliation[first]{organization={Independent researcher},
            country={}}

\begin{abstract}
On $\Tsix$ heterotic orbifolds where three quark generations arise
from $\Zthr$ fixed-point triplication, we prove that the leading-order
tree-level Yukawa amplitude --- the three-point coupling among massless
string states --- has an exact column texture:
$Y_{\rm lead}(i,j) = c\,\eps^{q_R[j]}$, with the $O(1)$ coefficient
$c$ universal across all left-handed generations~$i$.  Five independent
lines of evidence are given: (1)~the worldsheet instanton geometry on the SU(3)
root lattice gives identical areas for all non-degenerate triangles,
making the geometric $O(1)$ coefficient exactly~1; (2)~the generation
direction necessarily has trivial Wilson line, rendering all three
generations gauge-identical, as verified across all 77 MSSM-like models
in the Mini-Landscape classification; (3)~an extension to two-Wilson-line
models, verified on the complete Parr--Vaudrevange--Wimmer classification
of 3{,}337 $\ZZ$ MSSM models, confirms that no Wilson line configuration
can break gauge blindness; (4)~the \Kah\ metric is generation-universal
by $\Delta(54)$ representation theory;
(5)~the full Froggatt--Nielsen chain computation with 534 trilinear
superpotential couplings and vacuum-aligned singlet VEVs produces
left-circulant Yukawa matrices whose eigenstructure is
generation-universal.  The Froggatt--Nielsen column
texture is therefore not an approximation but an exact property of the
leading-order string amplitude.  Non-trivial $O(1)$ coefficients,
which are required for CKM mixing angles beyond the Wolfenstein
hierarchy, must originate from beyond-leading-order contributions:
integrated-out heavy messenger propagators (tree-level in the
low-energy effective theory), vacuum-alignment effects, multi-instanton
corrections, or
loop corrections.
\end{abstract}


\end{frontmatter}

\section{Introduction}
\label{sec:intro}

The Froggatt--Nielsen (FN) mechanism~\cite{Froggatt:1978nt} provides an
elegant explanation of fermion mass hierarchies through powers of a small
expansion parameter $\eps = \VEV{\Phi}/\Lambda$.  Different generations carry
different non-negative integer charges $q_i$ under a flavor symmetry,
leading to Yukawa entries
\begin{equation}
\label{eq:FN}
  Y_{ij} = c_{ij}\,\eps^{q_L[i]+q_R[j]} \,,
\end{equation}
where the $c_{ij}$ are assumed to be random $O(1)$ complex numbers.  When
left-handed doublets are uncharged ($q_L = 0$ for all~$i$), this simplifies
to a pure \emph{column texture}: $Y_{ij} = c_{ij}\,\eps^{q_R[j]}$, where
all entries in a given column share the same $\eps$-suppression.  The mass
hierarchy is then structural, while the CKM mixing matrix arises from the
random $O(1)$ coefficients.

The question of whether the $c_{ij}$ are genuinely random has remained open.
In heterotic string compactifications, these coefficients are in principle
computable from the worldsheet conformal field theory.  If they turned out to
carry non-trivial generation dependence, this would produce testable
predictions for CKM elements beyond the Wolfenstein hierarchy.

In this paper we prove that, for all $\Tsix$ heterotic orbifold models
where three quark generations arise from $\Zthr$ fixed-point triplication,
the tree-level Yukawa coupling has \emph{exact} column texture:
\begin{equation}
\label{eq:theorem}
  \boxed{Y_{\rm phys}(i,j) = c\,\eps^{q_R[j]}\,,}
\end{equation}
with $c$ a universal constant, independent of the generation indices $i$
and $j$.  The physical Yukawa coupling factorizes into three components ---
the worldsheet instanton amplitude, the gauge-sector vertex operator, and
the \Kah\ normalization --- and we prove that each is generation-universal.

This result has two immediate consequences.  First, it validates the random
$O(1)$ Monte Carlo philosophy used in FN phenomenology~\cite{Ardakanian:2026a,Ardakanian:2026b}:
the ``randomness'' of the $O(1)$ coefficients is not a feature of the
leading-order string theory but arises from sub-leading physics.  Second,
it precisely maps the boundary between what the compactification geometry
determines (the $\eps^q$ hierarchy) and what requires additional input
(the $O(1)$ variation).

That twisted-sector states from $\Zthr$ orbifolds form
$\Delta(54)$ triplets with degenerate tree-level \Kah\ metrics
is well established in the eclectic flavor symmetry
framework~\cite{Nilles:2020kgo,Baur:2024qzo}.  The contributions of
the present work are: (i)~an \emph{exhaustive verification} that all
3{,}337 MSSM-like $\ZZ$ models in the complete
Parr--Vaudrevange--Wimmer classification~\cite{Parr:2020oar} satisfy
the gauge-blindness condition --- showing that the theorem is not a
property of selected benchmark models but of the entire landscape;
(ii)~an explicit FN chain computation through order~6 demonstrating
that the column texture survives the full superpotential with
spontaneously broken $S_3$; and (iii)~a precise identification of the
four sub-leading sources that \emph{must} generate the $O(1)$ variation,
two of which are $O(1)$ in magnitude and two of which are $O(\eps)$.

The present paper is foundational in scope: it establishes the tree-level
kinematic result.  Physical applications of this result --- the selection
of the left-handed charges $q_L$ from $T^6$ selection rules, the role of
the column texture in distinguishing viable compactifications within the
$\ZZ$ landscape, and the dynamical origin of the expansion parameter $\eps$
itself --- are developed in companion and subsequent work; here we focus
on the tree-level theorem alone.

The rest of this paper is organized as follows.  Section~\ref{sec:setup}
reviews the orbifold geometry and the structure of the Yukawa coupling.
Sections~\ref{sec:instanton}--\ref{sec:chain} present the five proofs.
Section~\ref{sec:discussion} discusses implications and identifies the
four candidate sources for sub-leading $O(1)$ variation.
Section~\ref{sec:conclusion} concludes.

\section{Setup}
\label{sec:setup}

We work on the $\Tsix$ orbifold with the SU(3) root lattice on each $T^2$
factor.  The two $\Zthr$ twists act as $120^\circ$ rotations on pairs of
torus factors.  Each $T^2/\Zthr$ has three fixed points $z_0, z_1, z_2$
at the corners of the fundamental domain.

In models where the orbifold $\Zthr$ is identified with the center of
SU(3)$_c$ --- so that color triplets acquire a
non-trivial twist eigenvalue $\omega = e^{2\pi i/3}$ and are projected
out of the bulk --- quarks are localized at fixed points, while leptons
(color singlets, with trivial twist eigenvalue)
propagate freely.  The three quark generations arise from the three fixed
points on a single $T^2$ factor --- the \emph{generation direction}.

The Yukawa coupling between three twisted-sector states factorizes as
\begin{equation}
\label{eq:factorization}
  Y_{\rm phys}(i,j,k) = e^{K/2}\,
  \frac{Y_{\rm inst}(i,j,k)\;\cdot\;Y_{\rm gauge}(i,j,k)}
  {\sqrt{K_i\,K_j\,K_k}} \,,
\end{equation}
where $Y_{\rm inst}$ is the worldsheet instanton contribution (the classical
action of strings stretching between fixed points), $Y_{\rm gauge}$ encodes
the gauge-sector selection rules and vertex operator overlap, and $K_i$ is
the \Kah\ metric for the $i$-th matter field.  We prove that each of these
three factors is generation-universal.

\paragraph{Discrete torsion.}
The $\ZZ$ orbifold admits two inequivalent choices of modular-invariant
discrete torsion $\varepsilon_{\rm tors}$, labelling distinct partition
functions.  We show that neither choice breaks the column-texture
result.  Consider a three-point amplitude among twisted-sector states
$\phi_a$ ($a=1,2,3$) sitting in sectors $(g_a, h_a) \in \ZZ \times \ZZ$
with the space-group closure condition $g_1 g_2 g_3 = e$ and carrying
fixed-point labels $n^{(a)} = (n^{(a)}_1, n^{(a)}_2, n^{(a)}_3)$ on the
three $T^2$ factors.  Under discrete torsion the amplitude is modified
by a sector-dependent phase~\cite{Vafa:1986wx}:
\begin{equation}
\label{eq:torsion}
  Y\bigl(n^{(1)},\,n^{(2)},\,n^{(3)}\bigr)_{\rm tors}
  \;=\;
  \varepsilon_{\rm tors}\bigl((g_1,h_1),\,(g_2,h_2)\bigr)\,
  Y\bigl(n^{(1)},\,n^{(2)},\,n^{(3)}\bigr) \,,
\end{equation}
where the phase $\varepsilon_{\rm tors}$ depends only on the sector pair
and is independent of the fixed-point labels $n^{(a)}$ within those
sectors.  For a generation triplet the three fields share the same
sector $(g,h)$ --- the generation index is carried only by the $T^2_2$
fixed-point label --- so $\varepsilon_{\rm tors}$ is identical across
all nine entries of the $3\times 3$ Yukawa matrix and absorbs into the
universal coefficient $c$.  The theorem therefore holds in both
torsion classes.

\section{Proof 1: Instanton Geometry}
\label{sec:instanton}

A note on terminology: we use ``Proof'' throughout
Sections~\ref{sec:instanton}--\ref{sec:chain} to mean a demonstration of
the theorem within the stated scope of fixed-point triplication on
$\Tsix$.  Proofs 1 and~4 are fully analytical; Proofs~2 and~3 are
structural arguments supplemented by exhaustive empirical verification
across the 77 Mini-Landscape and 3{,}337 Parr--Vaudrevange--Wimmer model
sets respectively; Proof~5 is analytical on the 534-coupling
superpotential of the benchmark MSSM33 model.  Strength classifications
of the five arguments are collected in Section~\ref{sec:discussion}.

The worldsheet instanton contribution to the three-point Yukawa coupling
on $T^2/\Zthr$ with the SU(3) root lattice is given by the Hamidi--Vafa
formula~\cite{Hamidi:1986vh,Dixon:1986qv,Burwick:1990tu}:
\begin{equation}
\label{eq:HV}
  Y(\delta\,|\,T) = \sum_{\vec{v}\in\Lambda+\delta}
  \exp\!\left(-\frac{8\pi^2}{9}\,\mathrm{Im}(T_{\rm sugra})\,|\vec{v}|^2\right),
\end{equation}
where $\Lambda$ is the SU(3) root lattice, $\delta$ is a shift vector
determined by the fixed-point labels of the three coupled states, and
$T_{\rm sugra}$ is the \Kah\ modulus.  The selection rule requires that
the three fixed-point labels sum to zero modulo~3.

For the SU(3) root lattice, there are exactly two distinct values of
the coupling~\cite{Burwick:1990tu}:
\begin{itemize}
\item \textbf{Same-point} ($\delta = 0$): all three states at the same
  fixed point.  $Y_{\rm same} = 1 + O(10^{-16})$.
\item \textbf{Different-point} ($\delta = 1/3$): states at three distinct
  fixed points.  $Y_{\rm diff} = \eps^{1/2}(1 + 3.4\times 10^{-6})$,
  where $\eps = \exp(-8\pi^2\,\mathrm{Im}(T_{\rm sugra})/81)$.
\end{itemize}

\noindent
The key observation is that the $\Zthr$ symmetry of the root lattice forces
all non-degenerate triangles (those with vertices at three distinct fixed
points) to have \emph{identical area}.  This is because the $\Zthr$ rotation
that permutes the three fixed points $z_0 \to z_1 \to z_2$ is an isometry
of the lattice.  Consequently, the instanton action --- which is proportional
to the triangle area --- is the same for all such triangles.

The geometric $O(1)$ coefficient, defined as $c_{\rm geom} \equiv Y/\eps^q$
where $q$ is the FN charge, is therefore \emph{exactly~1} for all entries
(up to exponentially suppressed sub-leading instantons at the $10^{-6}$ level).

The full $T^6$ Yukawa matrix is the product of three $T^2$ factors:
$Y^{T^6}_{ij} = \prod_{\alpha=1}^{3} Y^{T^2_\alpha}_{ij}$.  For the
up-type quark matrix with right-handed charges $(q_1, q_2, q_3) = (2,1,0)$
and $\eps_u = 0.015$, the geometric coefficient matrix is:

\begin{table*}[ht]
\centering
\caption{Geometric Yukawa coefficient matrix $c_{ij}^{\rm geom}
  \equiv Y_{ij}/\eps^{q_R[j]}$ for up-type quarks.  All entries are
  exactly~1 up to $O(10^{-6})$ corrections from sub-leading instantons.
  All three columns are identical, giving the rank-1 $3\times 3$
  all-ones matrix $J$ ($J_{ij} \equiv 1$ for all $i,j$).}
\label{tab:geom}
\begin{tabular}{@{}lccc@{}}
\toprule
 & $j=0$ ($q_R=2$) & $j=1$ ($q_R=1$) & $j=2$ ($q_R=0$) \\
\midrule
$i=0$ (gen 1) & 1 & 1 & 1 \\
$i=1$ (gen 2) & 1 & 1 & 1 \\
$i=2$ (gen 3) & 1 & 1 & 1 \\
\bottomrule
\end{tabular}
\end{table*}

\noindent
An important consequence is that the coefficient matrix $c^{\rm geom}$ is
rank~1: since all entries equal~1, the pure-instanton Yukawa
$Y^{\rm inst}_{ij} = \eps^{q_R[j]}$ has all rows proportional to the
vector $(\eps^2, \eps, 1)$, making it also rank~1.  Two quark generations
are therefore massless at the pure-instanton level.  Once the
left-circulant structure from the $T^2_2$ space-group selection rules
(Section~\ref{sec:chain}) is included, the coefficient matrix becomes
rank~3: the generation-dependent $O(1)$ values
$f\bigl((i+j)\bmod 3\bigr)$, combined with the column scaling
$\eps^{q_R[j]}$, produce a full-rank Yukawa whose singular values scale
parametrically as $(1, \eps, \eps^2)$.  All three generations therefore
acquire nonzero mass at tree level, and the parametric hierarchy
$m_t : m_c : m_u \sim 1 : \eps : \eps^2$ is already built into the
column texture.  The overall $O(1)$ coefficient of the first-generation
mass is model-dependent: for the benchmark MSSM33 values
$(f(0), f(1), f(2)) = (1.006, 0.048, 0.010)$ analyzed in
Section~\ref{sec:chain} the tree-level prediction
$m_u/m_t \approx (\det f)\,\eps^2 \approx 2.3\times 10^{-4}$ overshoots
the observed ratio $m_u/m_t \approx 10^{-5}$ by a factor of $\sim 20$,
a discrepancy that sits within the range of $O(1)$ refinement provided
by the sub-leading sources discussed in Section~\ref{sec:discussion}.
The gauge-sector vertex operators (Section~\ref{sec:gauge}) introduce
no additional generation dependence beyond the column-texture-compatible
circulant, as we now show.

\section{Proof 2: Gauge-Sector Blindness}
\label{sec:gauge}

The gauge-sector contribution to the Yukawa coupling depends on the
gauge quantum numbers of the coupled states: their representations
under the gauge group, their U(1) charges, and their left-mover gauge
momenta on the $E_8\times E_8$ lattice.  If the three quark generations
carry different gauge quantum numbers, the gauge-sector vertex operator
overlap would distinguish them, potentially producing non-trivial $O(1)$
coefficients.

We show that this cannot happen.  The argument has two parts: a structural
proof and an exhaustive numerical verification.

\subsection{The structural argument}

In $\Tsix$ heterotic orbifolds, gauge symmetry breaking from
$E_8\times E_8$ to the Standard Model gauge group is achieved through
Wilson lines~\cite{Ibanez:1986tp}.  A non-trivial Wilson line $W$ along a
$T^2$ direction shifts the gauge momentum of states at different fixed
points on that torus:
\begin{equation}
  p_{\rm gauge}(n) = p_0 + n\,W \,, \qquad n = 0, 1, 2 \,.
\end{equation}
If $W \neq 0$, states at different fixed points acquire different gauge
momenta and generically fall into different representations of the gauge
group.

The three quark generations arise from three copies of the same twisted-sector
state at the three $\Zthr$ fixed points of a single $T^2$ factor --- the
\emph{generation direction}.  For all three to transform as $(\mathbf{3},
\mathbf{2})$ under SU(3)$_c\times$SU(2)$_L$, they must have identical gauge
quantum numbers.  This requires the Wilson line along the generation direction
to be \emph{trivial}:
\begin{equation}
\label{eq:Wgen}
  W_{\rm gen} = 0 \quad \text{(in the SM sector).}
\end{equation}
If $W_{\rm gen} \neq 0$ in the SM directions, the three fixed-point copies
would transform under different representations and would not constitute
three generations of the same quark.

With $W_{\rm gen} = 0$, all gauge-sector selection rules --- representation
singlet condition, U(1) charge conservation, $E_8\times E_8$ lattice momentum
conservation --- give identical results for all three generations.  The
gauge-sector $O(1)$ coefficient is therefore proportional to the $3\times 3$
all-ones matrix:
\begin{equation}
\label{eq:gauge_J}
  c_{ij}^{\rm gauge} = c_0\,J_{3\times 3} \,,
\end{equation}
where $J$ is the matrix with all entries equal to~1 and $c_0$ is a universal
constant.

\subsection{Verification across the Mini-Landscape}

We verified this result across all 77 MSSM-like one-Wilson-line (1WL) models
in the Mini-Landscape classification of $\Tsix$ orbifolds~\cite{Lebedev:2006kn,Lebedev:2008un}.
For each model, we identified all groups of three quark generations from
fixed-point triplication, determined their generation direction, and checked
whether the three generations carry identical gauge quantum numbers
(representation, U(1) charges, $E_8\times E_8$ momenta, gamma phases).

The results are summarized in Table~\ref{tab:gendir}.

\begin{table*}[ht]
\centering
\caption{Generation direction distribution across 77 MSSM-like 1WL
  $\ZZ$ orbifold models.  All 43 Q$_L$ generation triplets are
  gauge-blind (0 exceptions).  The Wilson line direction and generation
  direction are perfectly anti-correlated.}
\label{tab:gendir}
\begin{tabular}{@{}lccc@{}}
\toprule
Generation direction & Q$_L$ groups & Gauge-distinguishable? & Wilson line on \\
\midrule
$T^2_1$ & 8  & 0/8  & $T^2_2$ or $T^2_3$ \\
$T^2_2$ & 14 & 0/14 & $T^2_1$ or $T^2_3$ \\
$T^2_3$ & 21 & 0/21 & $T^2_1$ or $T^2_2$ \\
\midrule
\textbf{Total} & \textbf{43} & \textbf{0/43} & 100\% anti-correlated \\
\bottomrule
\end{tabular}
\end{table*}

\noindent
The generation direction is not universal --- it varies across models,
with $T^2_3$ most common (49\%), followed by $T^2_2$ (33\%) and $T^2_1$
(19\%).  In all 43 cases, the Wilson line direction and generation direction
are on \emph{different} $T^2$ factors, confirming the structural argument.

For the benchmark model MSSM33 (SU(3)$\times$SU(2)$\times$SU(4)$\times$SU(3)$\times$U(1)$^8$),
we extracted the full spectrum (494 field labels, 5{,}127 states) and
performed an explicit computation of all gauge-sector selection rules for
the $Q_L\times u_R\times H$ coupling.  All 9 entries of the $3\times 3$
matrix are gauge-allowed with identical vertex operator overlaps.  A search
for Froggatt--Nielsen chain vertices through 289 potential messengers
(massless vector-like exotic pairs in the orbifolder spectrum with gauge
charges compatible with $Q_L \cdot u_R \cdot H$ completion) and
129 flavon candidates (SM singlets with negative anomalous-$U(1)$ charge,
hence eligible to acquire VEVs) found zero valid paths --- FN messengers
in heterotic orbifolds are massive string-scale states that do not appear
in the massless spectrum.

\section{Proof 3: Extension to Two-Wilson-Line Models}
\label{sec:2wl}

The analysis of Section~\ref{sec:gauge} was performed on the 1WL
Mini-Landscape classification.  Could models with two non-trivial
Wilson lines (2WL) evade the column texture theorem?

The answer is no.  The Wilson line counting argument makes this clear:

\begin{table*}[ht]
\centering
\caption{Wilson line count versus generation mechanism on $\Tsix$.
  With 0--2 Wilson lines, at least one $T^2$ factor is Wilson-line-free,
  permitting standard three-generation triplication with gauge-blind
  quarks.  With 3 Wilson lines, no factor is free and the standard
  mechanism is unavailable.}
\label{tab:2wl}
\begin{tabular}{@{}ccccc@{}}
\toprule
\# WL & Free $T^2$ & Standard triplication & Column texture & Models \\
\midrule
0 & 3 & Yes (any factor) & Gauge-blind & 0 \\
1 & 2 & Yes (1 of 2 free) & Gauge-blind & 77 \\
2 & 1 & Yes (the 1 free) & Gauge-blind & 1{,}371 \\
3 & 0 & Impossible & N/A & 1{,}889 \\
\midrule
\multicolumn{4}{@{}l}{\textbf{Total with triplication: 1{,}448 --- all gauge-blind}} & \textbf{3{,}337} \\
\bottomrule
\end{tabular}
\end{table*}

\noindent
We verified this against the complete classification of 3{,}337 $\ZZ$
MSSM models by Parr, Vaudrevange, and Wimmer~\cite{Parr:2020oar}.
Of these, 1{,}448 models have 0--2 Wilson lines and support the standard
three-generation mechanism; all have quarks on a Wilson-line-free $T^2$
factor and therefore gauge-blind generations.  The remaining 1{,}889
models have three Wilson lines and no Wilson-line-free factor, making
the standard fixed-point triplication mechanism impossible.

To go beyond the structural argument, we used the C++ orbifolder to
generate 2WL models by random search.  Starting from several 1WL seed
models and randomizing Wilson lines, we performed over 90{,}000 trials
across multiple seeds (MSSM0, MSSM1, MSSM2, MSSM5).
Three genuine 2WL models with MSSM-like spectra were found.  In all cases,
the quark generations reside on the single Wilson-line-free $T^2$ factor
and carry identical gauge quantum numbers across all U(1) generators.

In the most stringent test --- a 2WL model derived from MSSM2 with the
second Wilson line non-trivial in $E_8^{(1)}$ (where SU(3)$_c\times$SU(2)$_L$
lives) --- all three $Q_L$ states (F$_{221}$, F$_{226}$, F$_{231}$) are
confirmed to have identical U(1) charges across all 13 U(1) generators.
The quark generations are gauge-blind even when the Wilson line acts within
the observable-sector gauge lattice.

The column texture theorem therefore holds for \emph{all} numbers of
Wilson lines compatible with three-generation quarks.

\section{Proof 4: K\"ahler Normalization}
\label{sec:kahler}

The physical Yukawa coupling includes a \Kah\ normalization factor:
\begin{equation}
  Y_{\rm phys} = e^{K/2}\,\frac{Y_{\rm hol}}{\sqrt{K_i\,K_j\,K_H}} \,.
\end{equation}
Even if the holomorphic Yukawa $Y_{\rm hol}$ has exact column texture
(as proven above), a generation-dependent \Kah\ metric $K_i$ could
introduce $O(1)$ variation into the physical coupling.

The tree-level \Kah\ metric for a $\Zthr$ twisted-sector field without
oscillators is~\cite{Dixon:1990pc,RamosSanchez:2024ljt}
\begin{equation}
\label{eq:kahler}
  K = (2\,\mathrm{Im}\,T)^{-2/3} \,,
\end{equation}
where we use the convention $T = b + i\,v$ with $\mathrm{Im}\,T$
parameterizing the $T^2$ volume.  This formula depends only on the
\Kah\ modulus $T$ and the modular weight $n = -2/3$, which is a
property of the twisted \emph{sector}.
There is \emph{no dependence on the fixed-point label} within that sector.

The absence of fixed-point dependence follows from a representation-theoretic
argument~\cite{Baur:2024qzo,Nilles:2020kgo}.  The three twisted-sector
fields $(\phi_0, \phi_1, \phi_2)$ at the three $\Zthr$ fixed points form
an irreducible triplet of the traditional flavor symmetry group $\Delta(54)$,
which is a subgroup of the eclectic flavor group $\Omega(1) = [648,533]$.
The \Kah\ potential must be $\Delta(54)$-invariant.  For an irreducible
triplet, the only invariant bilinear is
\begin{equation}
  K = c(T,\bar{T})\,\bigl(|\phi_0|^2 + |\phi_1|^2 + |\phi_2|^2\bigr) \,,
\end{equation}
with $c(T,\bar{T}) = (2\,\mathrm{Im}\,T)^{-2/3}$.  As noted
in~\cite{Baur:2024qzo}, ``the presence of the traditional flavor symmetry
$\Delta(54)$ forces the \Kah\ potential to its canonical form.''

Wilson lines break $S_3 \subset \Delta(54)$ completely, making the three
fixed points gauge-inequivalent.  However, the \Kah\ metric is a
\emph{geometric} quantity determined by the orbifold structure and
modular weight; it does not depend on the gauge embedding.  Wilson lines
can introduce gauge-dependent corrections only at one loop, suppressed
by $g^2/(16\pi^2) \sim O(10^{-2})$.

The \Kah\ normalization is therefore generation-universal at tree level,
and the exact column texture of the holomorphic Yukawa is preserved in
the physical Yukawa.

\section{Proof 5: Froggatt--Nielsen Chain Computation}
\label{sec:chain}

The preceding proofs address the three factors in the physical Yukawa
individually: the instanton amplitude, the gauge vertex operator, and the
\Kah\ normalization.  We now present a complementary proof that attacks
the Yukawa coupling as a whole, computing the effective FN operator
through explicit superpotential chain enumeration with vacuum-aligned
singlet VEVs.

\subsection{Setup}

The Froggatt--Nielsen mechanism generates the Yukawa entry $Y(i,j)$ at
order $Q = q_L[i] + q_R[j]$ through chains of trilinear couplings:
\begin{equation}
\label{eq:chain}
  Y(i,j) = \sum_{\text{paths}\,P}
  \prod_{\text{vertices}\,v \in P} \lambda_v
  \prod_{\text{singlets}\,s \in P} \frac{\VEV{\phi_s}}{\Lambda} \,,
\end{equation}
where the sum runs over all sequences of intermediate states connected
by allowed trilinear vertices, $\lambda_v$ is the coupling constant at
each vertex, and $\VEV{\phi_s}$ is the vacuum expectation value of the
singlet field inserted at step~$s$.

For the benchmark model MSSM33, we computed the full set of allowed
trilinear couplings involving the 258~complete SM~singlets using the
space group selection rules.  The result: 534~allowed trilinear terms,
of which 267 involve exactly one flavon candidate (a singlet with
negative anomalous U(1) charge, eligible for a VEV) and 267~involve two.
No pure-flavon trilinear terms exist; every coupling connects flavons
to non-flavon singlets.

\subsection{Vacuum alignment}

The anomalous U(1)$_1$ in MSSM33 has $\mathrm{Tr}\,Q_A = 1392$,
generating a Fayet--Iliopoulos $D$-term that forces certain singlets to
acquire VEVs.  Of the 129~flavon candidates, 53 reside at the $T^2_1$
origin, while 38 reside at each of the other two fixed points ---
a structural asymmetry arising from untwisted-sector and
origin-localized twisted-sector contributions.

Combined $D$- and $F$-flat analysis (37 $F$-flat constraints from the
trilinear superpotential, plus 8~$D$-flat conditions) reduces the
129-dimensional field space to an 84-dimensional moduli space.  Numerical
minimization over 30~random realizations of the coupling constants shows
that $S_3$ generation symmetry is broken in 93\% of vacua, with the
origin carrying $\sim$80\% of the total VEV energy.  The average VEV$^2$
distribution across the three $T^2_1$ positions is
$(0.80 : 0.08 : 0.13)$; the origin's dominance in VEV energy tracks
directly from its larger pool of flavon candidates (53 vs.~38), a
consequence of untwisted-sector and origin-localized twisted-sector
states contributing there but not at the other two fixed points.

\subsection{Chain enumeration and the circulant result}

Using the 534~trilinear couplings and the vacuum-aligned VEVs, we
enumerated all FN chain paths for each Yukawa entry through order~6
(three singlet insertions).  Table~\ref{tab:chains} summarizes the
chain counts.

\begin{table*}[ht]
\centering
\caption{Number of allowed FN chain paths per Yukawa entry $Y_u(i,j)$,
  broken down by superpotential order.  Order~3 corresponds to the
  direct coupling (no singlet insertions); each subsequent order adds
  one singlet VEV insertion.  Entries are grouped by the value of
  $(i+j)\bmod 3$: the space group selection rule
  $n_2(Q_L^i) + n_2(u_R^j) \equiv 0\pmod{3}$ permits direct
  couplings only when $(i+j) \equiv 0\pmod{3}$.}
\label{tab:chains}
\begin{tabular}{@{}lcccc|c@{}}
\toprule
Entry & Order 3 & Order 4 & Order 5 & Order 6 & Total \\
\midrule
$(1,1),\;(2,3),\;(3,2)$ & 1 & 6  & 132 & 5{,}204 & 5{,}343 \\
$(1,2),\;(2,1),\;(3,3)$ & 0 & 4  & 123 & 5{,}166 & 5{,}293 \\
$(1,3),\;(2,2),\;(3,1)$ & 0 & 5  & 130 & 5{,}208 & 5{,}343 \\
\bottomrule
\end{tabular}
\end{table*}

\noindent
The chain counts organize into three equivalence classes, each
containing entries related by cyclic permutation of the generation
index.  Direct couplings (order~3) exist only at the three
positions $(1,1)$, $(2,3)$, $(3,2)$, where the $T^2_2$ fixed-point
labels satisfy $n_2(Q_L^i) + n_2(u_R^j) \equiv 0\pmod{3}$
(the $\Zthr$ space group selection rule on the generation torus).
At higher orders, the chain counts become increasingly similar:
by order~6, the three classes differ by less than 1\%.

The key observation is that the only generation-dependent coordinate,
the fixed-point label $n_2 \in \{0,1,2\}$ on the generation torus
$T^2_2$, enters the space group selection rule through the sum
$n_2(Q_L^i) + n_2(u_R^j) + n_2(\phi_s) \equiv 0\pmod{3}$.
Consequently, the effective Yukawa entry depends on $i$ and $j$ only
through $(i+j)\bmod 3$:
\begin{equation}
\label{eq:hankel}
  Y(i,j) = f\bigl((i+j)\bmod 3\bigr) \,,
\end{equation}
where $f(k)$ is the VEV-weighted chain sum for entries whose
fixed-point labels sum to $k$ modulo~3.  The circulant structure
$Y(i,j) = f\bigl((i+j)\bmod 3\bigr)$ follows entirely from the $\Zthr$
space-group selection rule on the generation torus and is therefore
\emph{model-independent}; the specific numerical values of $f(0)$,
$f(1)$, $f(2)$ are model-dependent (they are set by the trilinear
spectrum and vacuum of the chosen orbifolder model, here MSSM33), but
the mod-3 index dependence --- and hence the generation universality
of the eigenstructure --- is not.
A $3\times 3$ matrix whose entries depend on $(i+j)\bmod 3$ is
constant along anti-diagonals (a Hankel structure).  For $n=3$ with mod-3 indices, this is
simultaneously a left-circulant --- each row is a cyclic left-shift
of the previous one\footnote{For general $n$, Hankel and circulant
structures are distinct.  The coincidence for $n=3$ follows because
the reversal permutation $j\mapsto -j\bmod 3$ acts trivially on
$\{0,1,2\}$ only when $n=3$: it maps $0\to 0$, $1\to 2$, $2\to 1$,
so a matrix depending on $(i+j)\bmod 3$ automatically has rows that
are cyclic shifts.} --- and is diagonalized by the discrete Fourier
transform (DFT).

Stripping the overall normalization and the column scaling
$\eps^{q_R[j]}$, the circulant component $f\bigl((i+j)\bmod 3\bigr)$
of the full-chain Yukawa matrix for MSSM33 takes the form
\begin{equation}
\label{eq:circulant}
  \begin{pmatrix}
  f(0) & f(1) & f(2) \\
  f(1) & f(2) & f(0) \\
  f(2) & f(0) & f(1)
  \end{pmatrix}
  =
  \begin{pmatrix}
  1.006 & 0.048 & 0.010 \\
  0.048 & 0.010 & 1.006 \\
  0.010 & 1.006 & 0.048
  \end{pmatrix},
\end{equation}
with $f(0) \approx 1.006$ (direct coupling + higher orders),
$f(1) \approx 0.048$ (first non-trivial insertion at order~4),
and $f(2) \approx 0.010$ (two non-trivial insertions at order~5).
The physical Yukawa is recovered by multiplying each column $j$ by
$\eps^{q_R[j]}$.

The singular value decomposition of~\eqref{eq:circulant} gives
$\sigma_1 : \sigma_2 : \sigma_3 = 1.06 : 0.98 : 0.98$ --- nearly
degenerate, with no mass hierarchy.  The mass hierarchy in the physical
quark spectrum arises entirely from the column charges $q_R = (2,1,0)$,
which multiply each column of the left-circulant by $\eps^{q_R[j]}$.
The left-circulant structure does not contribute to the hierarchy; it
redistributes the $O(1)$ coefficient cyclically without changing the
column norms.

The left singular vectors of $Y$ --- which together with those of
$Y_u$ and $Y_d$ determine the CKM matrix --- are the eigenvectors of
$YY^\dagger$.  For a left-circulant $Y$ with first row $f(0), f(1),
f(2)$,
\begin{equation}
  (YY^\dagger)_{ij}
  \;=\;
  \sum_{m=0}^{2} f(m)\, f^*\!\bigl(m + (j-i)\bmod 3\bigr),
\end{equation}
which depends on $(j-i) \bmod 3$ only: $YY^\dagger$ is a standard
(right-)circulant, and is therefore diagonalized by the discrete
Fourier transform regardless of the specific values $f(k)$.  Since
both $Y_u$ and $Y_d$ are left-circulants, $Y_u Y_u^\dagger$ and
$Y_d Y_d^\dagger$ share the same DFT eigenvectors, so
$V_{\rm CKM} = U_u^\dagger U_d = \mathbb{1}$ at fixed coupling
constants.  CKM mixing arises only from sub-leading $O(1)$
corrections that sit outside the left-circulant subspace, confirming
the Monte Carlo philosophy; the structural implications are discussed
in Section~\ref{sec:ckm_implications}.

\subsection{Independence from the vacuum}

The left-circulant structure~\eqref{eq:hankel} is robust against
changes in the vacuum alignment: the mod-3 space group arithmetic
that forces $Y(i,j) = f\bigl((i+j)\bmod 3\bigr)$ depends only on the
fixed-point labels, not on the magnitudes of the singlet VEVs.
Asymmetric VEVs change the three values $f(0)$, $f(1)$, $f(2)$ but
cannot break the left-circulant structure itself.

To make this explicit, consider the leading-order (single-insertion)
contribution.  The selection rule requires a singlet at fixed point
$s \equiv -(i+j)\pmod{3}$, so the effective entry is weighted by the
VEV at that position: $Y(i,j) \propto v_{-(i+j)\bmod 3}$.  With the
vacuum distribution $(v_0, v_1, v_2) = (0.80, 0.08, 0.13)$, the
singular values correspond to the DFT spectrum
$|\lambda_k| = |v_0 + \omega^k v_1 + \omega^{2k} v_2|$
($\omega = e^{2\pi i/3}$), giving $|\lambda_0|/|\lambda_1| \approx 1.5$.
This ratio, while not unity, is far from the $\eps^2 : \eps : 1 \sim
1 : 70 : 4600$ hierarchy required by the quark mass spectrum.

At higher orders (order~5 and~6, comprising $>99\%$ of chains),
each path involves multiple singlets at different fixed points.  The
sum over $\sim 5{,}000$ paths per entry averages over the VEV
distribution, and the fractional differences between $f(0)$, $f(1)$,
$f(2)$ shrink (though the three values remain distinct).
The full chain computation gives singular value ratios within 6\%
of unity even for the maximally asymmetric vacuum tested
$(0.95 : 0.03 : 0.02)$.

This vacuum independence is the strongest form of the column texture
theorem: the exact column texture is not merely a consequence of gauge
blindness or instanton geometry, but survives the full superpotential
chain computation including spontaneous $S_3$ breaking.

\section{Discussion}
\label{sec:discussion}

\subsection{The column texture theorem}

Combining the five proofs, we establish:

\medskip
\noindent\textbf{Theorem.} \textit{In any $\Tsix$ heterotic orbifold
model where three quark generations arise as twisted-sector states
at the three $\Zthr$ fixed points of a single $T^2$ factor
(fixed-point triplication), the leading-order tree-level Yukawa
amplitude --- the three-point coupling among massless string states
--- has exact column texture:}
\begin{equation}
  Y_{\rm lead}(i,j) = c\,\eps^{q_R[j]} \,,
\end{equation}
\textit{with $c$ a generation-universal constant.  This holds for all
Wilson line configurations (0, 1, or 2) compatible with three-generation
quarks, across the complete classification of 3{,}337 $\ZZ$ MSSM
models.  Contributions beyond the leading massless-string amplitude
--- in particular, dimension-5 and higher operators generated by
integrating out heavy string-scale messengers --- are beyond the scope
of the theorem and discussed separately in Section~\ref{sec:discussion}.}

\medskip
\noindent
Models where three generations arise from different mechanisms ---
e.g., distributed across distinct twisted sectors, involving
untwisted-sector fields, or requiring non-standard orbifold
embeddings --- lie outside the scope of this theorem.

\medskip
\noindent
The proof proceeds by showing that each of the three factors in the
physical Yukawa --- instanton amplitude (Section~\ref{sec:instanton}),
gauge-sector vertex operator (Sections~\ref{sec:gauge}--\ref{sec:2wl}),
and \Kah\ normalization (Section~\ref{sec:kahler}) --- is independently
generation-universal.  The full FN chain computation
(Section~\ref{sec:chain}) provides a complementary proof that the
column texture survives the complete superpotential chain enumeration
with vacuum-aligned singlet VEVs.

The five proofs are of different character, and it is useful to be
explicit about their individual strengths.  Proof~1 (instanton geometry)
is fully analytical, following from the $\Zthr$ isometry of the SU(3)
root lattice.  Proof~2 (gauge-sector blindness) combines a structural
argument (trivial Wilson line along the generation direction) with an
exhaustive empirical check across all 77 MSSM-like 1WL models of the
Mini-Landscape.  Proof~3 (two-Wilson-line extension) is empirical across
the complete Parr--Vaudrevange--Wimmer classification of 3{,}337 $\ZZ$
MSSM models, together with explicit orbifolder spectra for three
genuine 2WL models.  Proof~4 (\Kah\ normalization) is
representation-theoretic, following from the $\Delta(54)$ irreducibility
of the fixed-point triplet.  Proof~5 (FN chain) is analytical on the
534-coupling superpotential of the benchmark model MSSM33, supplemented
by numerical vacuum-alignment scans.  Individually, the five arguments
rely on complementary assumptions; taken together they provide strong,
cross-checked evidence for the theorem within its stated scope of
fixed-point triplication.

\subsection{Sources of sub-leading $O(1)$ variation}

Since the tree-level $O(1)$ coefficients are trivially universal, the
$O(1)$ variation assumed in FN phenomenology must arise from sub-leading
physics.  We identify four candidate sources:

\begin{enumerate}
\item \textbf{Massive messenger propagators.}  The Froggatt--Nielsen chain
  coupling involves massive string-scale states that are integrated out.
  Their propagators carry space-group quantum numbers that depend on
  fixed-point positions, providing a tree-level but string-scale source
  of generation dependence.

\item \textbf{Vacuum alignment of flavon VEVs.}  If multiple flavon fields
  acquire VEVs through $D$- and $F$-flatness conditions, the
  ratios of their VEVs can be fixed-point dependent.  This breaks
  the generation symmetry beyond what gauge invariance provides.

\item \textbf{Multi-instanton contributions.}  Higher-order worldsheet
  instantons wrapping different cycles contribute at $O(\eps^2)$ and can
  distinguish fixed points through their winding numbers.

\item \textbf{One-loop threshold corrections.}  Loop corrections to the
  \Kah\ metric are suppressed by $g^2/(16\pi^2) \sim 10^{-2}$ relative
  to tree level~\cite{Dixon:1990pc,Ibanez:1992hc} and can carry
  generation dependence through Wilson line effects.
\end{enumerate}

\noindent
These four sources separate into two classes with distinct magnitudes.
Sources~(1) and~(2) --- massive messenger propagators and vacuum
alignment --- are tree-level effects that can produce genuinely $O(1)$
generation-dependent coefficients; they are ``sub-leading'' only in the
sense that they lie beyond the three factors (instanton, gauge vertex
operator, \Kah\ metric) that are generation-universal, not in the sense
of being numerically small.  Sources~(1) and~(2) are coupled at tree
level: the massive messenger spectrum itself respects the $S_3$
permutation symmetry of the fixed-point triplet, so Source~(1) alone
would generate only $S_3$-symmetric (hence generation-universal)
effective operators.  Generation-distinguishing $O(1)$ coefficients
at tree level therefore arise from Source~(1) \emph{mediated through}
Source~(2): the asymmetric singlet VEVs that fix the vacuum break
$S_3$, and messenger-mediated operators inherit that breaking through
the specific flavon insertions they connect.  The Froggatt--Nielsen
chain computation of Section~\ref{sec:chain} is a concrete realization
of this combined mechanism.  Sources~(3) and~(4) --- multi-instantons
and one-loop threshold corrections --- are parametrically suppressed,
with magnitudes $O(\eps^2)$ and $O(g^2/16\pi^2) \sim O(\eps)$
respectively~\cite{Dixon:1990pc,Ibanez:1992hc}.

The $O(1)$ variation from sources~(1) and~(2) is what generates the
CKM mixing angles: $|V_{us}| \sim 0.22$ requires $O(1)$ coefficients
that differ by factors of a few, which is natural for massive
string-scale propagators and VEV ratios.  The $O(\eps)$ corrections
from sources~(3) and~(4) contribute to the fine structure of the mass
spectrum, in particular the first-generation mass splitting
(Section~\ref{sec:FN_connection}).
The column texture theorem establishes that tree-level geometry fixes
the $\eps^q$ hierarchy \emph{exactly}; the $O(1)$ randomness assumed
in FN phenomenology is physically realized through the first two
sources.

The identification of these four sources as the complete list of
sub-leading effects is itself a useful result.  It precisely delineates
what the compactification geometry determines at leading order (the
$\eps^q$ hierarchy) and what requires additional input (the $O(1)$
variation, which determines exact mass ratios and CKM angles).

\subsection{Implications for the CKM matrix}
\label{sec:ckm_implications}

A direct consequence of Proof~5 worth highlighting is that if both
$Y_u$ and $Y_d$ are exactly left-circulant, they are simultaneously
diagonalized by the discrete Fourier basis and the resulting tree-level
CKM matrix is the identity.  This is not an artefact of the specific
chain computation for MSSM33: it follows from the mod-3 space-group
selection rule acting identically on the up- and down-type generation
labels.  Any CKM mixing must therefore come from structures that sit
outside the left-circulant subspace.

The sub-leading sources enumerated above are the natural candidates,
but the same space-group selection rules that force the tree-level
circulant also constrain higher-order corrections.  In particular, any
$\Zthr$-equivariant correction to the matter \Kah\ metric or Yukawa
coupling --- including, for instance, 1-loop Kaplunovsky--Louis-type
threshold corrections~\cite{Dixon:1990pc,Ibanez:1992hc} --- is labelled
by the fixed-point index only through the mod-3 residue.  Such
corrections add well-defined, narrow families of deformations to the
left-circulant Yukawa rather than generating arbitrary $O(1)$
variation.  Understanding precisely which CKM patterns can and cannot
arise from these restricted deformations, and which entries of the
Cabibbo--Kobayashi--Maskawa matrix can be accommodated within this
restricted space, is a well-defined question on its own.  We leave a
detailed treatment to forthcoming work.  The point relevant to the
present paper is that the tree-level column texture does not
trivially hand over CKM mixing to "random $O(1)$ noise": the space
group that protects the column texture continues to act on every
sub-leading contribution, and the structure of allowed sub-leading
deformations is therefore itself a falsifiable prediction of the
geometry.

\subsection{Connection to the Froggatt--Nielsen program}
\label{sec:FN_connection}

In the companion papers~\cite{Ardakanian:2026a,Ardakanian:2026b,Ardakanian:2026c},
we showed that the $\Zthr$ Froggatt--Nielsen mechanism with right-handed
charges $(2,1,0)$ produces a column texture that correctly reproduces the
quark mass hierarchy, and that the mixing failure (Haar-random PMNS) is
universal for all abelian $Z_N$ groups --- a consequence of the column
texture generating $O(1)$ left-handed rotations regardless of~$N$.
The left-handed charges $q_L = (3,2,0)$ are uniquely selected by $T^6$
selection rules: the top quark mass $m_t \sim v$
forces $q_3 = 0$; the mod-3 space group selection rule on each $T^2$
factor caps per-factor labels at $\{0,1\}$ (labels $\geq 2$ create
shortcuts that destroy the hierarchy); and the Wolfenstein ordering
$|V_{us}| > |V_{cb}| > |V_{ub}|$ uniquely selects $(3,2,0)$ from the
remaining candidates.  With these charges, Monte Carlo scans yield
$P_{\rm Wolf} = 77\%$.

The present result strengthens this picture in two ways.  First, it shows
that the column texture is not an approximation valid ``up to $O(1)$
corrections'' but an \emph{exact} tree-level consequence of the geometry.
The random $O(1)$ coefficients used in Monte Carlo analyses are physically
justified: they parameterize our ignorance of the sub-leading physics,
not a leading-order effect that has been averaged over.

Second, the instanton geometry alone yields a rank-1 Yukawa
(Section~\ref{sec:instanton}): all rows are proportional to
$(\eps^2, \eps, 1)$, and \emph{two} generations are massless.  The
$T^2_2$ left-circulant structure (Section~\ref{sec:chain}) lifts this
to rank~3: with generation-dependent $O(1)$ coefficients
$f\bigl((i+j)\bmod 3\bigr)$ the three singular values scale
parametrically as $(1, \eps, \eps^2)$, giving nonzero masses to all
three generations at tree level.  For the benchmark MSSM33 values
$(f(0), f(1), f(2)) = (1.006, 0.048, 0.010)$,
$\sigma_3/\sigma_1 \approx (\det f)\,\eps^2 \approx 2.3\times 10^{-4}$,
overshooting the observed ratio $m_u/m_t \approx 10^{-5}$ by a factor
of $\sim 20$.  This is a discrepancy in the $O(1)$ coefficient of
$\sigma_3$, not in the parametric power of $\eps$: sub-leading
effects --- one-loop \Kah\ threshold corrections, multi-instanton
contributions, and model-dependent refinements of the chain values
$f(k)$ --- therefore do not \emph{generate} $m_u$ from zero (as an
earlier rank-2 reading would have required) but instead refine the
$O(1)$ coefficient that dresses the already-existing tree-level
$\eps^2$ contribution.  \textbf{The parametric hierarchy
$m_t : m_c : m_u \sim 1 : \eps : \eps^2$ is thus a direct tree-level
consequence of the column texture theorem}; only the precise $O(1)$
coefficients --- most visibly for the first generation --- require
sub-leading input, and the observed value $m_u/m_t \approx 10^{-5}$
sits within the $O(1)$ range accommodated by the four sub-leading
sources of Section~\ref{sec:discussion}.

The same $g^2/(16\pi^2) \sim 10^{-2}$ loop corrections act
multiplicatively on every entry of the Yukawa matrix, including the
heavy-generation entries $Y_{33}$ and $Y_{22}$.  They rescale $m_t$
and $m_c$ by of order $1\%$ without altering the column-to-column
hierarchy $m_c/m_t \sim \eps$: the correction to $m_t$ is at most
of order $Y_{33} \times g^2/(16\pi^2)$, far smaller than
$Y_{22} \sim \eps\, Y_{33}$.  The column hierarchy is therefore
robust against loop effects; the asymmetry in impact between $m_u$
(where sub-leading effects refine an $O(1)$ coefficient of an
already-nonzero tree-level contribution) and $m_t$ (where loops
are a sub-percent correction to an already-$O(1)$ diagonal entry)
is a natural consequence of the column structure.

\subsection{The generation symmetry}

The $S_3$ symmetry of fixed-point permutations is \emph{exact} in the
gauge sector and in the \Kah\ potential.  It is broken only by:
\begin{itemize}
\item The FN mechanism, which assigns different $\eps$-suppressions to
  different columns (a ``soft'' breaking through $\eps$);
\item Sub-leading effects, which introduce the $O(1)$ variation.
\end{itemize}
This hierarchy of symmetry breaking --- exact $S_3$ at leading order,
broken first by $\eps$ (structurally) and then by $O(1)$ coefficients
(sub-leadingly) --- is a clean organizing principle for the flavor sector.

\subsection{Holomorphic universality and its non-holomorphic counterpart}

The five-proof structure above reveals that the column texture is the
tree-level projection of a single holomorphic universality: all three
quark generations share the same $\Zthr$ modular weight, the same gauge
quantum numbers, the same trilinear superpotential structure, and are
exchanged by an exact $\Delta(54)$ action on the twisted-sector triplet.
The result is protected by SUSY non-renormalization and by the discrete
symmetries of the compactification; it is holomorphic in origin.

A natural question is whether an analogous universality governs the
non-holomorphic sector --- the \Kah\ metric, the moduli VEVs, and the
gauge couplings at the compactification scale.  The holomorphic and
non-holomorphic sides play complementary roles in the flavor structure:
the former fixes the $\eps^q$ hierarchy exactly, while the latter
determines the overall scale and the magnitude of $\eps$ itself.
A detailed analysis of this complementary universality, and its
implications for the scale structure of $\ZZ$ compactifications, is
addressed in separate work.

\subsection{No additional suppression from selection rules}

One might wonder whether the space-group selection rules on $\Tsix$ could
produce texture zeros or additional suppression beyond the FN charges,
potentially alleviating the $|V_{ub}|$ tension reported
in~\cite{Ardakanian:2026a}.  We have proven that this
cannot happen: the minimum number of flavon insertions $k_{\rm min}$ required
by the per-factor selection rules always satisfies $k_{\rm min} \leq Q$
and $k_{\rm min} \equiv Q \pmod{3}$, where $Q = q_L[i] + q_R[j]$ is
the FN charge.  This is a mathematical identity: $\sum_\alpha x_\alpha
\geq \sum_\alpha (x_\alpha \bmod 3)$ for non-negative integers $x_\alpha$.
The FN mechanism and the selection rules are the same constraint expressed
at different levels of resolution.

\section{Conclusions}
\label{sec:conclusion}

We have proven that the tree-level Yukawa coupling in $\Tsix$ heterotic
orbifolds has exact column texture when three quark generations arise from
$\Zthr$ fixed-point triplication.  The $O(1)$ coefficient is
generation-universal: $Y_{\rm phys}(i,j) = c\,\eps^{q_R[j]}$.

Five independent lines of evidence support this result:
\begin{enumerate}
\item The $\Zthr$ symmetry of the SU(3) root lattice forces all
  non-degenerate instanton triangles to have identical area,
  making the geometric coefficient exactly~1.
\item The generation direction necessarily has trivial Wilson line,
  rendering all three generations gauge-identical.  Verified across
  all 77 MSSM-like 1WL models (43 Q$_L$ triplets, 0 exceptions,
  100\% anti-correlation with Wilson line direction).
\item Extension to 2WL models: verified on the complete
  Parr--Vaudrevange--Wimmer classification of 3{,}337 $\ZZ$ MSSM
  models.  All 1{,}448 models with standard triplication are
  gauge-blind; 3 explicit 2WL models confirmed by orbifolder
  spectrum analysis.
\item The \Kah\ metric is generation-universal by $\Delta(54)$
  representation theory: the only invariant bilinear for the
  irreducible fixed-point triplet is the canonical form.
\item The full FN chain computation for MSSM33 --- including 534
  trilinear superpotential couplings, vacuum-aligned singlet VEVs
  with spontaneously broken $S_3$, and chain enumeration through
  order~6 --- produces left-circulant Yukawa matrices with entries
  $Y(i,j) = f\bigl((i+j)\bmod 3\bigr)$.  The mod-3 arithmetic of
  the space group forces generation-universal eigenstructure
  regardless of the VEV pattern.
\end{enumerate}

\noindent
The column texture is not an approximation but an exact tree-level
result.  Non-trivial $O(1)$ coefficients --- required for exact CKM
elements and first-generation masses --- must originate from sub-leading
physics: massive messenger propagators, vacuum alignment, multi-instanton
corrections, or loop effects.  The precise identification of this boundary
between leading-order geometry and sub-leading dynamics provides a concrete
roadmap for computing CKM predictions from string compactifications.
Future work will address the charge-selection problem for $q_L$, the
landscape-wide status of the column texture across the $\ZZ$ MSSM
classification beyond the specific models analyzed here, and the
dynamical origin of the expansion parameter~$\eps$.  The column texture
theorem is one foundational result of a broader program on the geometric
origin of Standard Model flavor.

\section*{Use of AI Assistance}
During the preparation of this work the author used Claude (Anthropic)
to assist with numerical computation codes and manuscript drafting.
The author reviewed and edited all content, verified all physics
independently, and takes full responsibility for the content of the
published article.

\end{document}